\documentclass[12pt]{article}
\usepackage{graphicx}
\usepackage{epsfig}
\usepackage{epstopdf}
\DeclareGraphicsExtensions{.pdf,.eps,.png,.jpg,.mps}
\usepackage{url}
\usepackage[implicit=false]{hyperref}

\usepackage{cite}

\def\lsim{\mathrel{\rlap{\lower3pt\hbox{\hskip0pt$\sim$}}
     \raise1pt\hbox{$<$}}}         
\def\gsim{\mathrel{\rlap{\lower4pt\hbox{\hskip1pt$\sim$}}
     \raise1pt\hbox{$>$}}}         

\usepackage{amsmath}
\usepackage{amsfonts}

\begin{document}
\begin{titlepage}

\centerline{\Large \bf Altcoin-Bitcoin Arbitrage}
\medskip

\centerline{Zura Kakushadze$^\S$$^\dag$\footnote{\, Zura Kakushadze, Ph.D., is the President of Quantigic$^\circledR$ Solutions LLC,
and a Full Professor at Free University of Tbilisi. Email: \href{mailto:zura@quantigic.com}{zura@quantigic.com}} and Willie Yu$^\sharp$\footnote{\, Willie Yu, Ph.D., is a Research Fellow at Duke-NUS Medical School. Email: \href{mailto:willie.yu@duke-nus.edu.sg}{willie.yu@duke-nus.edu.sg}}}
\bigskip

\centerline{\em $^\S$ Quantigic$^\circledR$ Solutions LLC}
\centerline{\em 1127 High Ridge Road \#135, Stamford, CT 06905\,\,\footnote{\, DISCLAIMER: This address is used by the corresponding author for no
purpose other than to indicate his professional affiliation as is customary in
publications. In particular, the contents of this paper
are not intended as an investment, legal, tax or any other such advice,
and in no way represent views of Quantigic$^\circledR$ Solutions LLC,
the website \url{www.quantigic.com} or any of their other affiliates.
}}
\centerline{\em $^\dag$ Free University of Tbilisi, Business School \& School of Physics}
\centerline{\em 240, David Agmashenebeli Alley, Tbilisi, 0159, Georgia}
\centerline{\em $^\sharp$ Centre for Computational Biology, Duke-NUS Medical School}
\centerline{\em 8 College Road, Singapore 169857}
\medskip
\centerline{(January 27, 2019)}

\bigskip
\medskip

\begin{abstract}
{}We give an algorithm and source code for a cryptoasset statistical arbitrage alpha based on a mean-reversion effect driven by the leading momentum factor in cryptoasset returns discussed in \url{https://ssrn.com/abstract=3245641}. Using empirical data, we identify the cross-section of cryptoassets for which this altcoin-Bitcoin arbitrage alpha is significant and discuss it in the context of liquidity considerations as well as its implications for cryptoasset trading.

\end{abstract}
\medskip

\end{titlepage}

\newpage

\section{Introduction}

{}There is a sizable proliferation of cryptoassets,\footnote{\, For the purposes of this note, cryptoassets include digital cryptography-based assets such as cryptocurrencies (e.g., Bitcoin), as well various other digital ``coins" and ``tokens" (minable as well as non-minable), that have data on \url{https://coinmarketcap.com}.} whose number according to \url{https://coinmarketcap.com} was 2,116 as of January 18, 2019. Just as with stocks (and other asset classes), there appear to be underlying common factors for cryptoasset returns, at least on shorter-horizons \cite{CryptoFM}.\footnote{\, \cite{CryptoFM} extends short-horizon equity factors \cite{4FM} to cryptoassets.} Thus, the leading common factor in the daily close-to-close cryptoasset returns is the prior day's momentum (``mom"), and on average the subsequent open-to-close return is negatively correlated with mom. So, there is a mean-reversion effect in daily cryptoasset returns.

{}Can we utilize this mean-reversion effect to construct a trading signal (alpha)? The mean-reversion here is cross-sectional, so such an alpha would (ideally) involve a sizable cross-section of cryptoassets. In the case of stocks one can construct a dollar-neutral mean-reversion strategy by going long a sizable number of stocks with the investment level $I_L$ and simultaneously shorting a sizable number of stocks with the investment level $I_S$, with $I_S = I_L$ (so we have dollar-neutrality), which is a standard long-short statistical arbitrage strategy. Alternatively, one can go long a sizable number of stocks with the investment level $I_L$ and simultaneously short an index futures with the investment level $I_S$ (again, $I_S = I_L$), e.g., S\&P 500 futures, in which case we have a dollar-neutral so-called S\&P 500 outperformance strategy.

{}We can attempt to do something similar with cryptoassets. However, shorting a sizable number of cryptoassets is not practicable. We can short Bitcoin futures\footnote{\, For a discussion on Bitcoin futures, see, e.g., \cite{BitFut}.} instead. The long position then can be in a sizable number of cryptoassets other than Bitcoin, i.e., altcoins. We then have a Bitcoin outperformance strategy, which we refer to as altcoin-Bitcoin arbitrage.\footnote{\, For some cryptoasset investment and trading related literature, see, e.g.,  \cite{Alessandretti2018}, \cite{Amjad2017}, \cite{Baek2014}, \cite{Bariviera2017}, \cite{Bouoiyour2016}, \cite{Bouri2017}, \cite{Brandvold2015}, \cite{Briere2015}, \cite{Cheah2015}, \cite{CheungRS2015}, \cite{Ciaian2015}, \cite{Colianni2015}, \cite{Donier2015}, \cite{Dyhrberg2015}, \cite{Eisl2015}, \cite{ElBahrawy}, \cite{Gajardo2018}, \cite{Garcia2015}, \cite{Georgoula2015}, \cite{Harvey2016}, \cite{JiangLiang2017}, \cite{KimKim2016}, \cite{Kristoufek2015}, \cite{Chuen2018}, \cite{Li2018}, \cite{LiewH2017}, \cite{LiewLB2018}, \cite{Nakano2018}, \cite{Ortisi2016}, \cite{Shah2014}, \cite{VanAlstyne2014}, \cite{Vo}, \cite{WangVergne2017}.} So, the idea is simple. We maintain a short Bitcoin position with the investment level $I_S$ (so we do not trade Bitcoin). The long position consists of a cross-section of altcoins, which changes daily based on their mom values, so we either establish or liquidate long altcoin positions, but never short them. We discuss some simple trading rules for altcoin positions in Section \ref{sec2}.

{}The next question is whether the alpha is ``tradable". The main considerations are transaction costs and liquidity. In this note we focus on liquidity considerations. Does this alpha work across essentially all cryptoassets, or is its performance correlated with their liquidity? We address this question empirically in Section \ref{sec3}. We briefly conclude in Section \ref{sec4}. Appendix \ref{app.A} gives R source code for building the portfolio and backtesting it.\footnote{\, The source code given in Appendix \ref{app.A} hereof is not written to be ``fancy" or optimized for speed or in any other way. Its sole purpose is to illustrate the algorithms described in the main text in a simple-to-understand fashion. Some important legalese is relegated to Appendix \ref{app.B}.} Tables and figures summarize our backtesting results.

\section{Alpha}\label{sec2}
\subsection{Setup and Data}

{}Cryptoassets trade continuously, 24/7.\footnote{\, Assuming no special circumstances such as trading halts.} Thus, ``open" on any given day means the price right after midnight (UTC time), while ``close" on any given day means the price right before midnight (UTC time), so the open on a given day is almost the same as the close of the previous day. All prices (open, close, high, low), volume and market cap are measured in dollars. The index $i=1,\dots,N$ labels $N$ different cryptoassets cross-sectionally, while the index $s=0,1,2,\dots$ labels the dates, with $s=0$ corresponding to the most recent date in the time series. So: $P^C_{is}$ (or, equivalently, $P^C_{i,s}$) is the close price for the cryptoasset labeled by $i$ on the day labeled by $s$; $P^O_{is}$ is the open price; $P^H_{is}$ is the high price; $P^L_{is}$ is the low price; $V_{is}$ is the daily dollar volume; $C_{is}$ is the market cap.\footnote{\, High, low and volume are measured between the open and the close of a given day.} All our data was freely downloaded (see below).

{}We define daily open-to-close log-returns (or ``continuously compounded" returns), which we use to define the mom factor as in \cite{CryptoFM} (see below):
\begin{equation}
 R_{is} = \ln\left({P^{C}_{is} / P^{O}_{is}}\right)
\end{equation}
For small values it is approximately the same as the standard (``single-period") return defined as
\begin{equation}
 {\widetilde R}_{is} = P^{C}_{is} / P^{O}_{is} - 1
\end{equation}
For computing portfolio P\&L, return, Sharpe ratio \cite{Sharpe1994}, etc., we use ${\widetilde R}_{is}$.

\subsection{Momentum}

{}There are various ways to define the momentum factor (mom). For our purposes here, we will define it as in \cite{CryptoFM}:
\begin{equation}\label{mom}
 \beta^{\rm{\scriptstyle{mom}}}_{is} = R_{i,s+1}
\end{equation}
This definition is 100\% out-of-sample: we use $\beta^{\rm{\scriptstyle{mom}}}_{is}$ for trading on the date labeled by $s$, and $\beta^{\rm{\scriptstyle{mom}}}_{is}$ is computed using quantities from the prior date labeled by $s+1$.

\subsection{Trading Signal}

{}The trading signal (alpha) $\alpha_{is}$ for altcoins is defined as follows:
\begin{eqnarray}\label{alpha}
 \alpha_{is} = \theta\left(-\beta^{\rm{\scriptstyle{mom}}}_{is}\right)
\end{eqnarray}
where the Heaviside function $\theta(x) = 1$ if $x > 0$, and $\theta(x) = 0$ if $x\leq 0$. So, we establish a new long altcoin position, or maintain an existing long altcoin position, if mom is negative. If mom is nonnegative, then we do not establish a new altcoin position, and liquidate an existing long altcoin position. All altcoin positions are long or null. Meanwhile, we continuously maintain a constant short Bitcoin position.\footnote{\, Technically, this should be short Bitcoin futures, but we assume a short Bitcoin position.}

\subsection{Altcoin Weights}

{}Let us assume that the constant short Bitcoin position has the investment level $I_S$. Let the total investment level of our long altcoin position be $I_L$. To have a dollar-neutral portfolio, we must set $I_L = I_S$. Let $H_{is}$ be the individual altcoin dollar holdings. Let us use the labels $i=2,\dots,N$ for altcoins, while $i=1$ will correspond to Bitcoin. We can define the altcoin weights as $w_{is} = H_{is} / I_L$. Then we have ($w_{is}\geq 0$):
\begin{equation}\label{norm}
 \sum_{i=2}^N w_{is} = 1
\end{equation}
The simplest choice for the weights is to have equal weights for all altcoins with nonzero signals:
\begin{eqnarray}\label{equal}
 &&w_{is} = {1\over n_s}~\alpha_{is}\\
 &&n_s = \sum_{i=2}^N \alpha_{is}
\end{eqnarray}
Other weighting schemes are possible, e.g., by suppressing the weights by volatility:
\begin{eqnarray}
 &&w_{is} = \gamma_s~ {\alpha_{is} \over \sigma_{is}}\\
 &&w_{is} = {\widetilde \gamma}_s~ {\alpha_{is}\left|\beta^{\rm{\scriptstyle{mom}}}_{is}\right|\over \sigma^2_{is}}\\
 &&\dots
\end{eqnarray}
Here $\sigma_{is}$ is historical volatility (e.g., the standard deviation of the returns $R_{is^\prime}$ computed over $d$ previous days $s^\prime = s+1,\dots,s+d$, or the hlv factor defined in \cite{CryptoFM}, etc.), while the normalization coefficients $\gamma_s$,${\widetilde \gamma}_s$ are fixed via Eq. (\ref{norm}). In our backtests (see below) we focus on equally weighted portfolios (\ref{equal}).

\section{Backtests}\label{sec3}
\subsection{Estimation Period and Universe}\label{sub.univ}

{}We downloaded\footnote{\, R source code for data downloads is given in Appendix A of \cite{CryptoFM}. This code still works with two minor tweaks. First, the line {\tt{\small u <- c(x[22:28])}} in the function {\tt{\small crypto.data()}} now reads {\tt{\small u <- c(x[22:27])}} due to a formatting change on \url{https://coinmarketcap.com}. Second, in the function {\tt{\small crypto.hist.prc()}}, right after the line {\tt{\small shared.write.table(x, file, T)}}, one should now include the following (or similar) line: {\tt{\small Sys.sleep(max(rnorm(1, 10, 2), 5))}} (which spaces downloads at random intervals). This is due to the apparent change at \url{https://coinmarketcap.com}, which averts ``rapid-fire" (i.e., continuous serial) downloads.\label{fn.data}} the data from \url{https://coinmarketcap.com} for all 2,116 cryptoassets as of January 19, 2019 (so the most recent date in the data is January 18, 2019), and 2,115 cryptoassets had downloadable data, albeit for many various fields were populated with ``?", which we converted into NAs. In our backtests (see below), we only kept cryptoassets with non-NA price (open, close, high, low), volume and market cap data, with an additional filter that no null volume was allowed either.\footnote{\, This is to avoid stale prices. Further, 2 cryptoassets had apparently ``artifact" stale prices during some periods, so they were also excluded from the corresponding backtests (see below).}

{}Cboe Bitcoin Futures (symbol XBT)\footnote{\, Which is the actual investment vehicle we implicitly assume for our short Bitcoin position.} started trading on December 10, 2017. So, technically speaking, backtesting the strategy before that time might not be particularly meaningful.\footnote{\, Especially considering the effect Bitcoin futures arguably had on Bitcoin (and other cryptoassets) -- see, e.g., \cite{BitFut}. For a nontechnical discussion, see, e.g., \cite{Kelleher}.} Still, although we primarily focus on the 1-year backtest (looking back from January 18, 2019), for comparison and completeness purposes we also run 2-, 3-, 4- and 5-year lookback backtests. For the 1-year backtest we have 417 cryptoassets with historical data, while for the 2-, 3-, 4- and 5-year backtests we have 121, 67, 44 and 13 cryptoassets, respectively.\footnote{\, These counts include Bitcoin. Also, these counts exclude the aforesaid 2 cryptoassets with apparently ``artifact" stale prices during some periods. Finally, in 2-year and longer backtests the cryptoasset ``Circuit of Value Coin" (symbol COVAL) is excluded as it had an extraordinarily large positive return in a short time period and including it would misleadingly ``rosy-up" the results.\label{fn.COVAL}} For the 1-year backtest we further break up the universe of the 416 altcoins (which together with Bitcoin make up the aforesaid 417 cryptoassets) into market cap tiers A,B,C,D,E,F. Thus, the alpha based on tier A on any given day goes long only the altcoins whose market cap on the previous day ranks 2 to 30 among all cryptoassets. Similarly, for tiers B,C,D,E,F the corresponding market cap rank ranges are: 31-60, 61-100, 101-200, 201-300, 301-417. In fact, based on our results (see below), running a backtest for the full universe of all 416 altcoins would obscure the liquidity effect (see below).

\subsection{Results}\label{sub.res}

\begin{figure}[ht]
\centering
\includegraphics[scale=1.0]{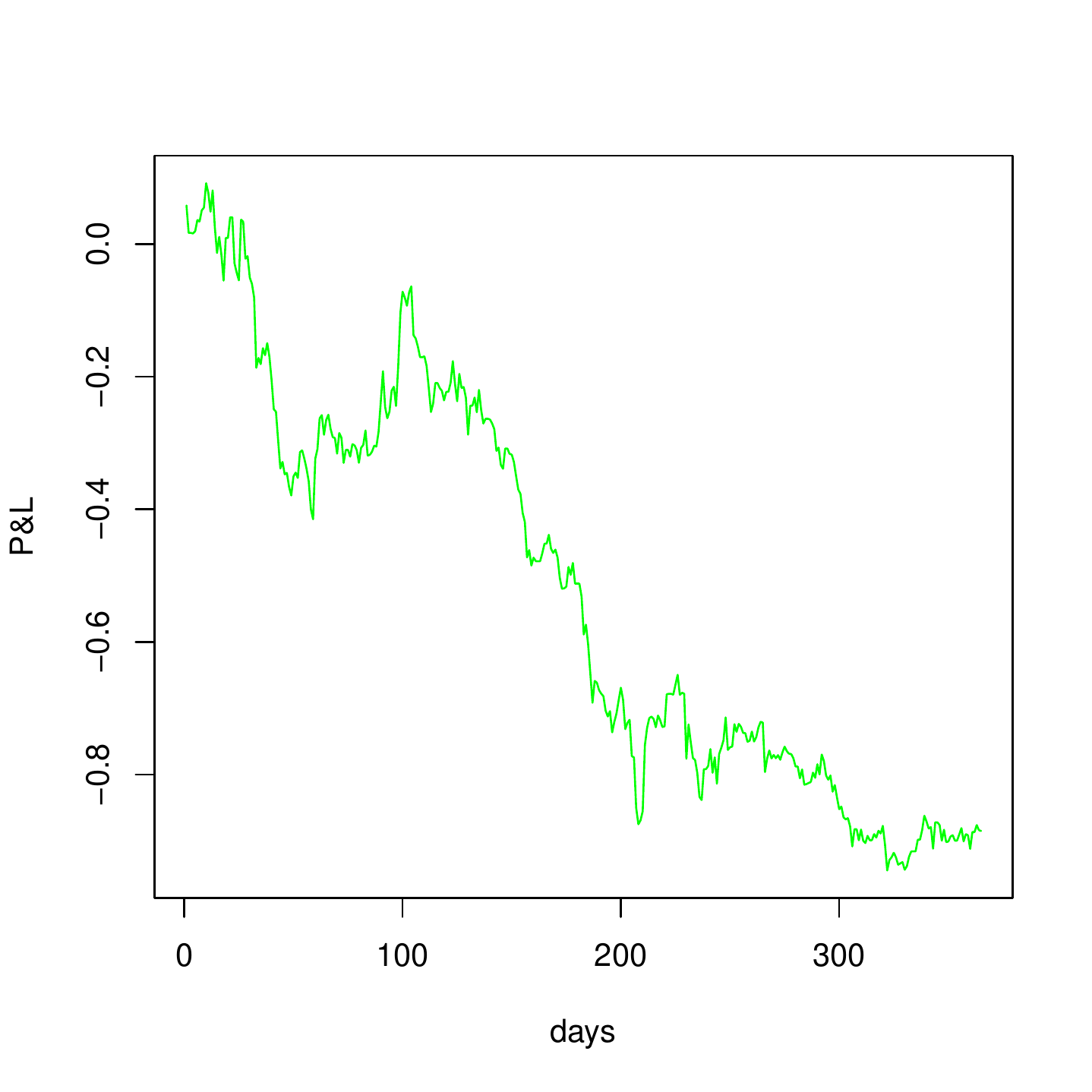}
\caption{P\&L (in the units where the short Bitcoin position is normalized to 1) for the lookback 1A portfolio (see Table \ref{table2}).}
\label{Figure1}
\end{figure}

\begin{figure}[ht]
\centering
\includegraphics[scale=1.0]{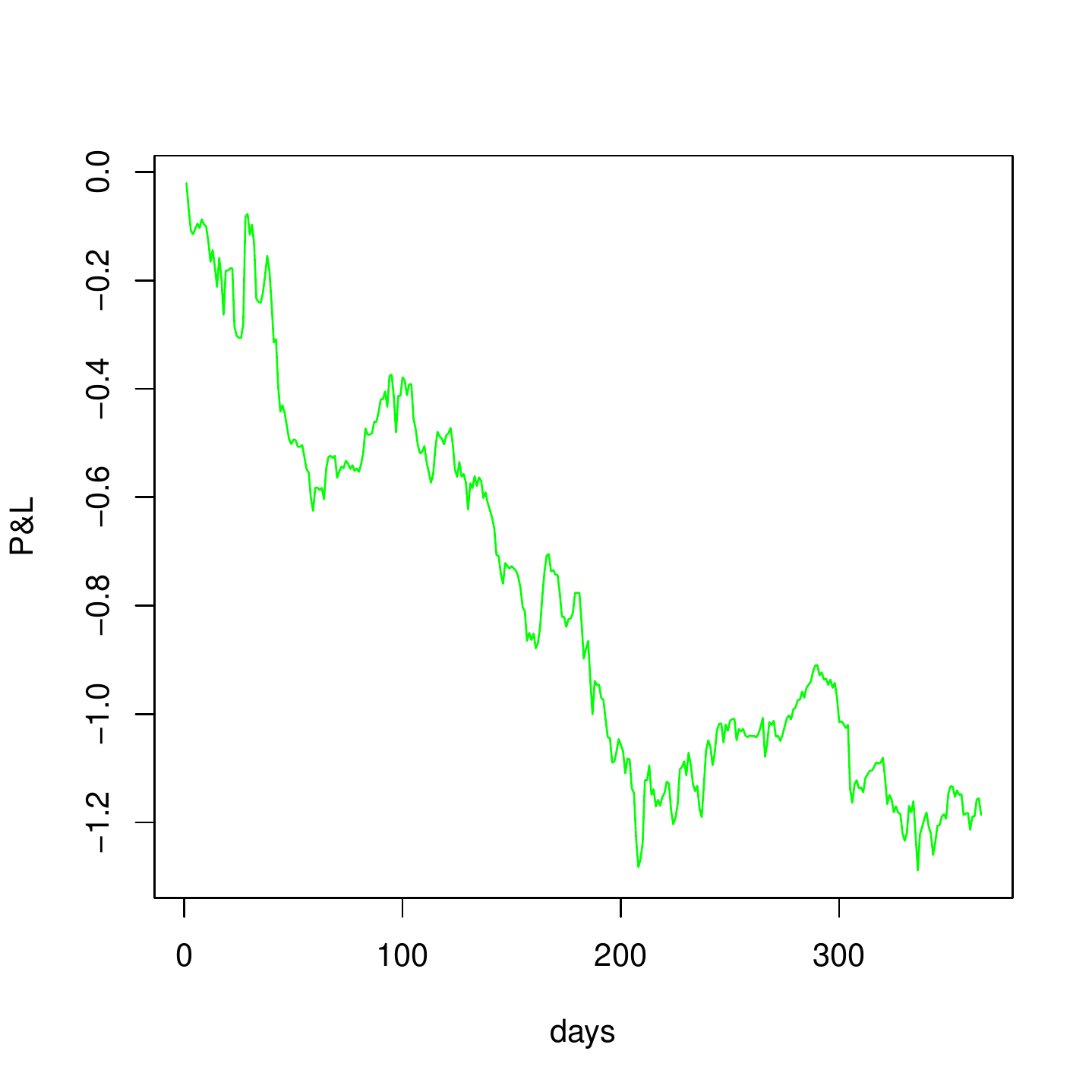}
\caption{P\&L (in the units where the short Bitcoin position is normalized to 1) for the lookback 1B portfolio (see Table \ref{table2}).}
\label{Figure2}
\end{figure}

\begin{figure}[ht]
\centering
\includegraphics[scale=1.0]{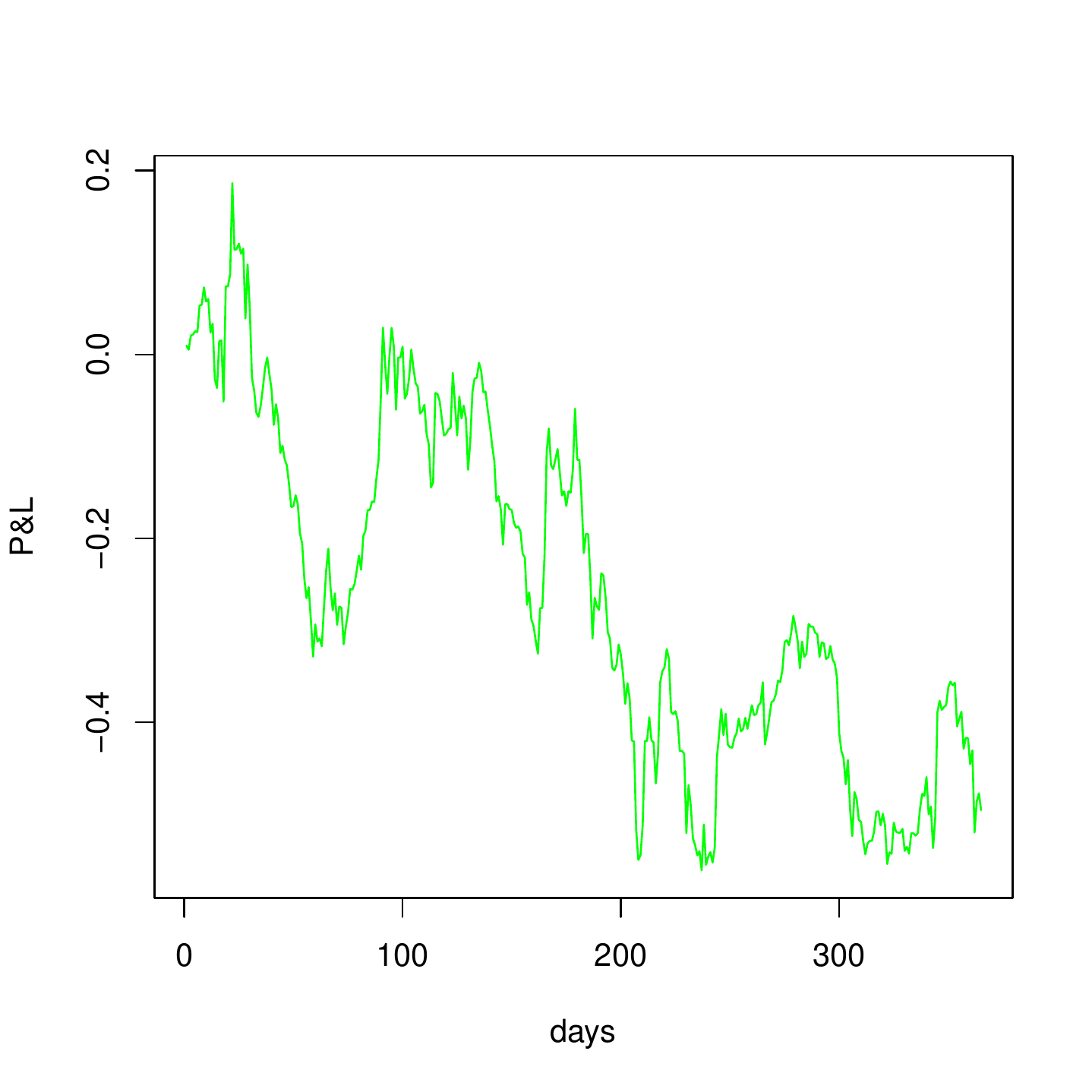}
\caption{P\&L (in the units where the short Bitcoin position is normalized to 1) for the lookback 1C portfolio (see Table \ref{table2}).}
\label{Figure3}
\end{figure}

\begin{figure}[ht]
\centering
\includegraphics[scale=1.0]{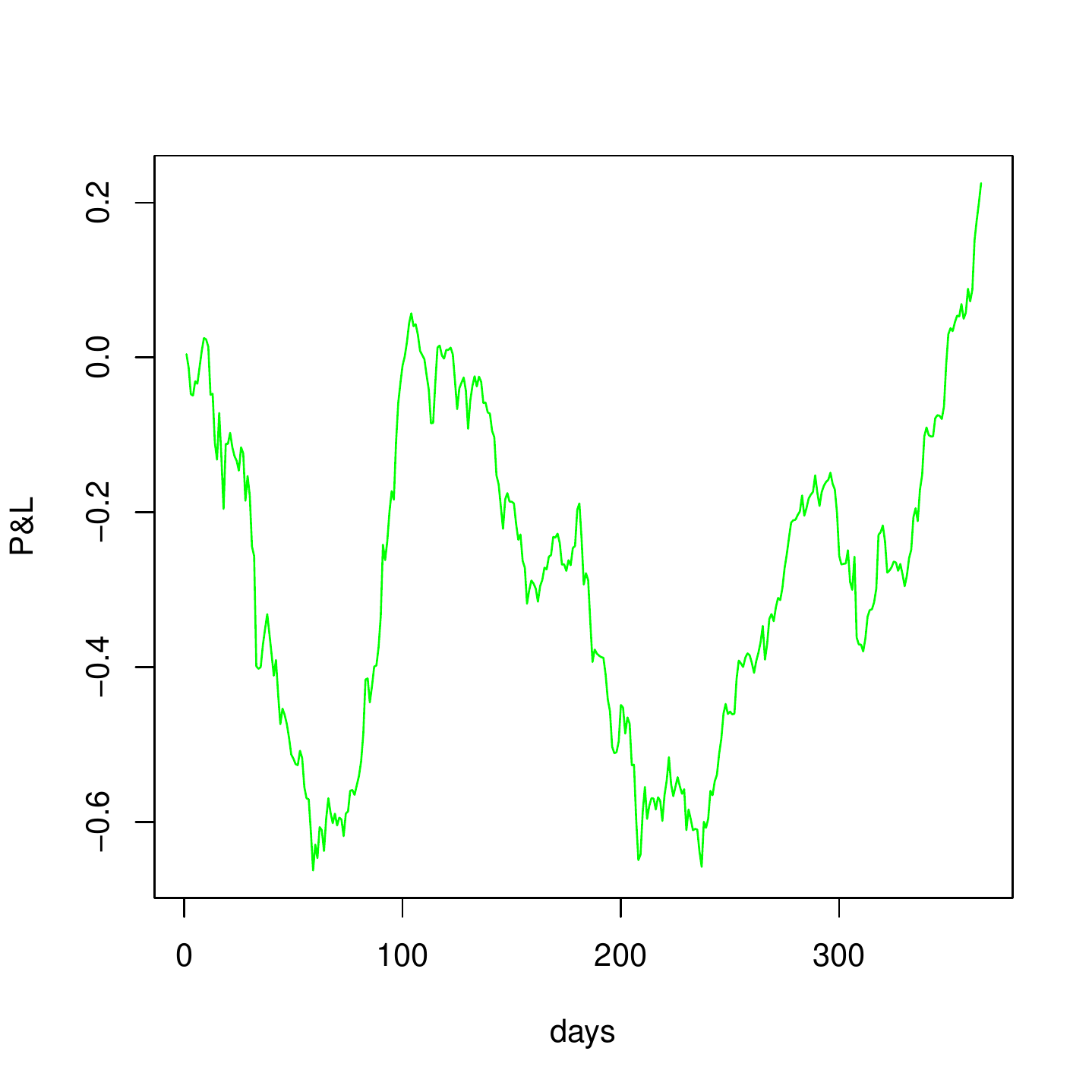}
\caption{P\&L (in the units where the short Bitcoin position is normalized to 1) for the lookback 1D portfolio (see Table \ref{table2}).}
\label{Figure4}
\end{figure}

\begin{figure}[ht]
\centering
\includegraphics[scale=1.0]{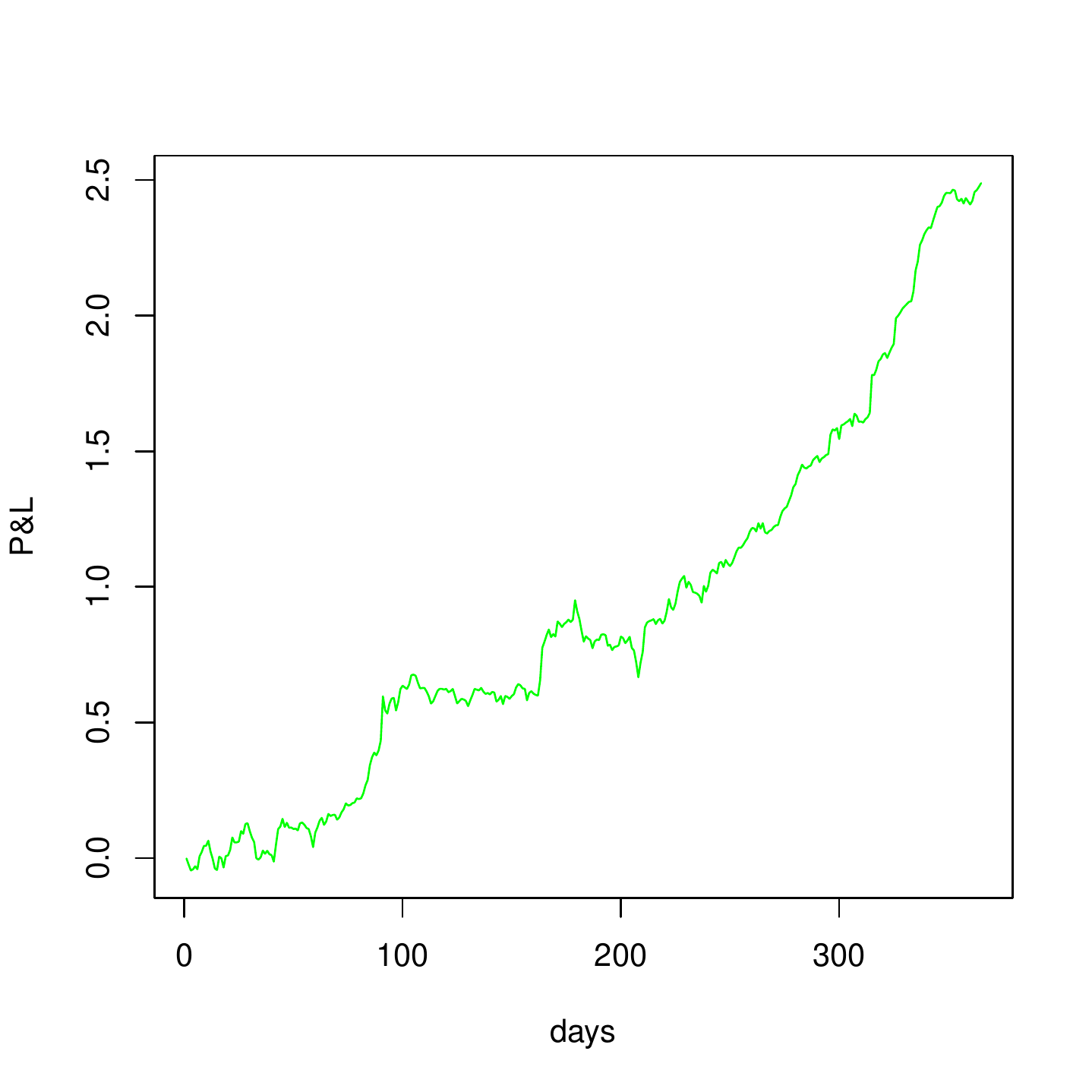}
\caption{P\&L (in the units where the short Bitcoin position is normalized to 1) for the lookback 1E portfolio (see Table \ref{table2}).}
\label{Figure5}
\end{figure}

\begin{figure}[ht]
\centering
\includegraphics[scale=1.0]{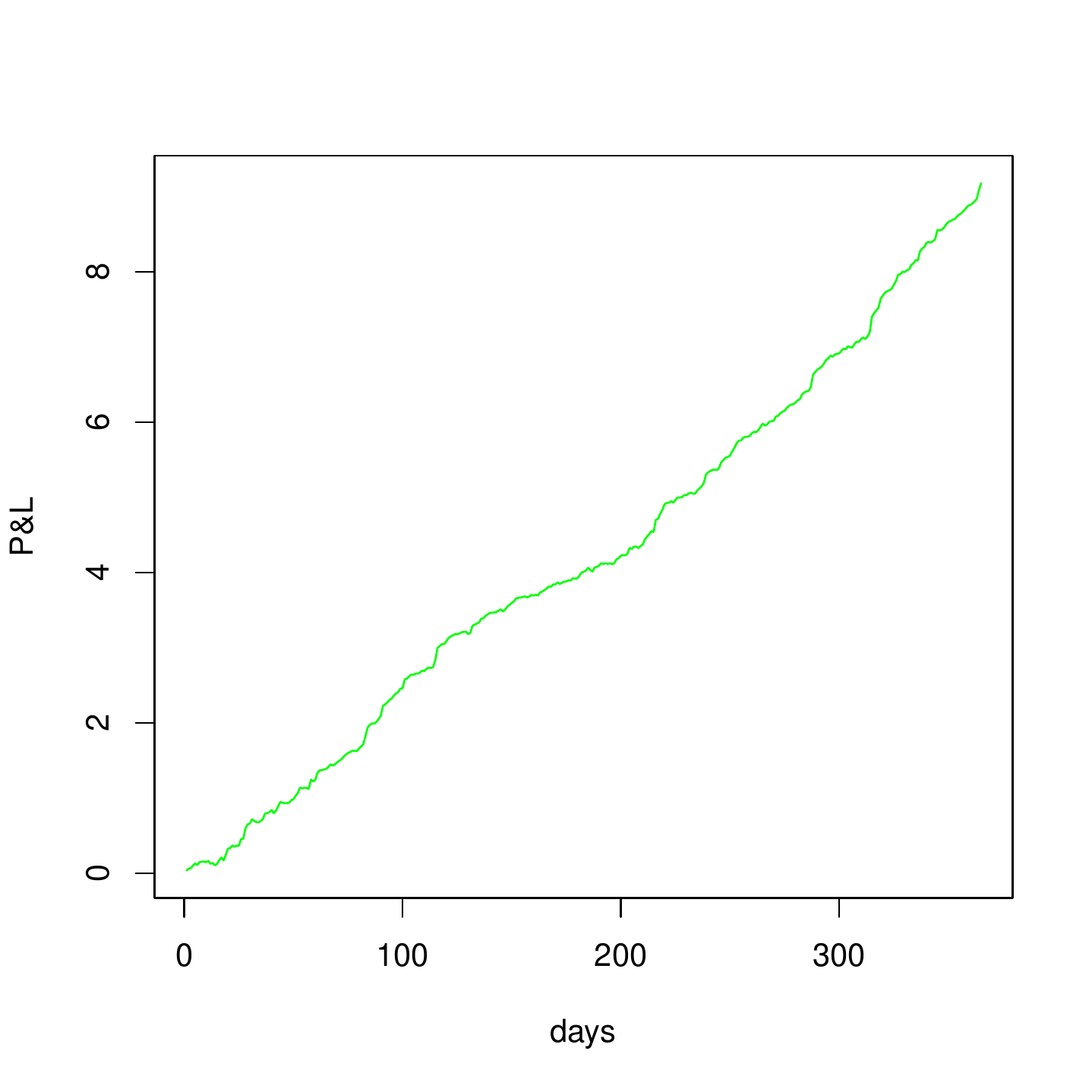}
\caption{P\&L (in the units where the short Bitcoin position is normalized to 1) for the lookback 1F portfolio (see Table \ref{table2}).}
\label{Figure6}
\end{figure}

\begin{figure}[ht]
\centering
\includegraphics[scale=1.0]{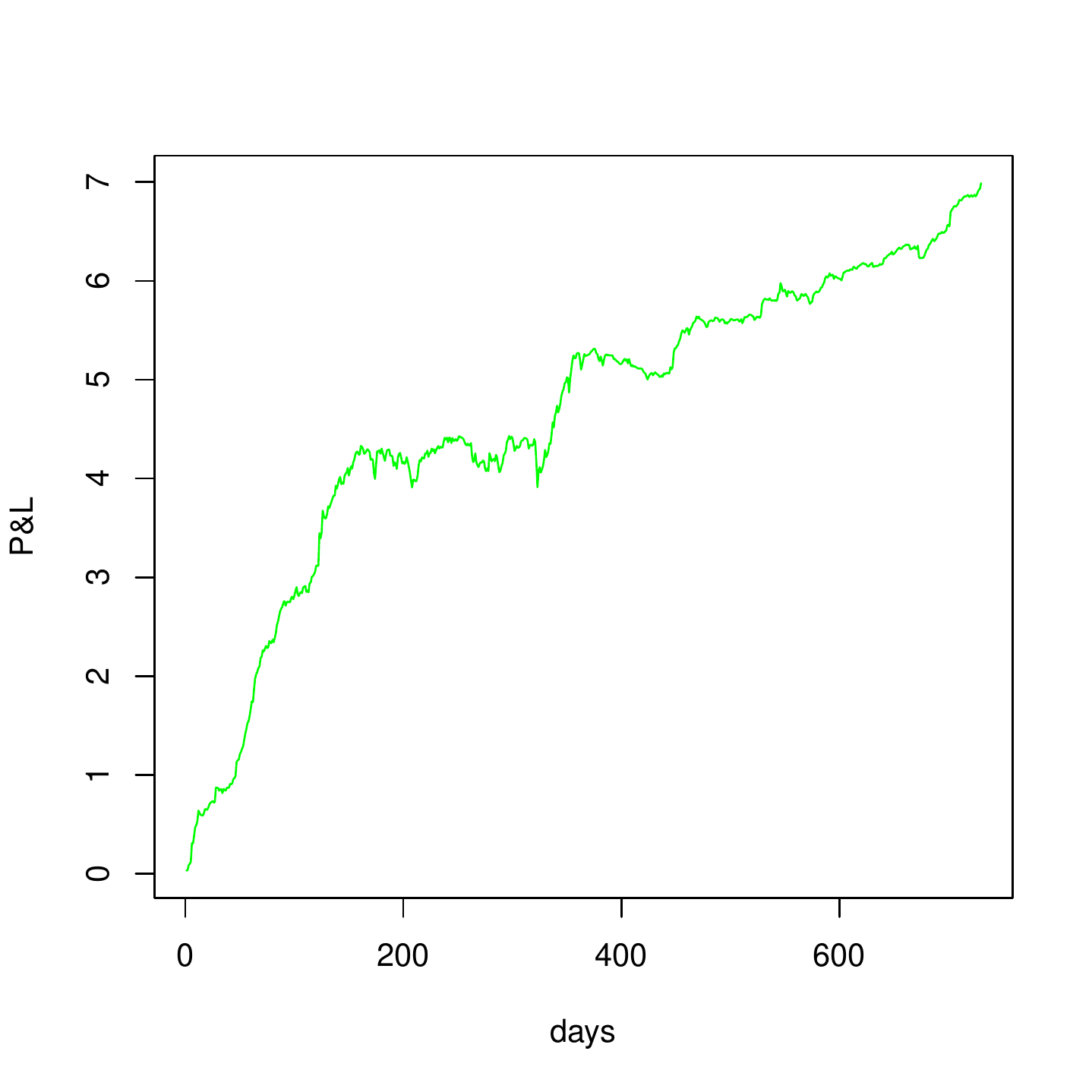}
\caption{P\&L (in the units where the short Bitcoin position is normalized to 1) for the lookback 2 portfolio (see Table \ref{table2}).}
\label{Figure7}
\end{figure}

\begin{figure}[ht]
\centering
\includegraphics[scale=1.0]{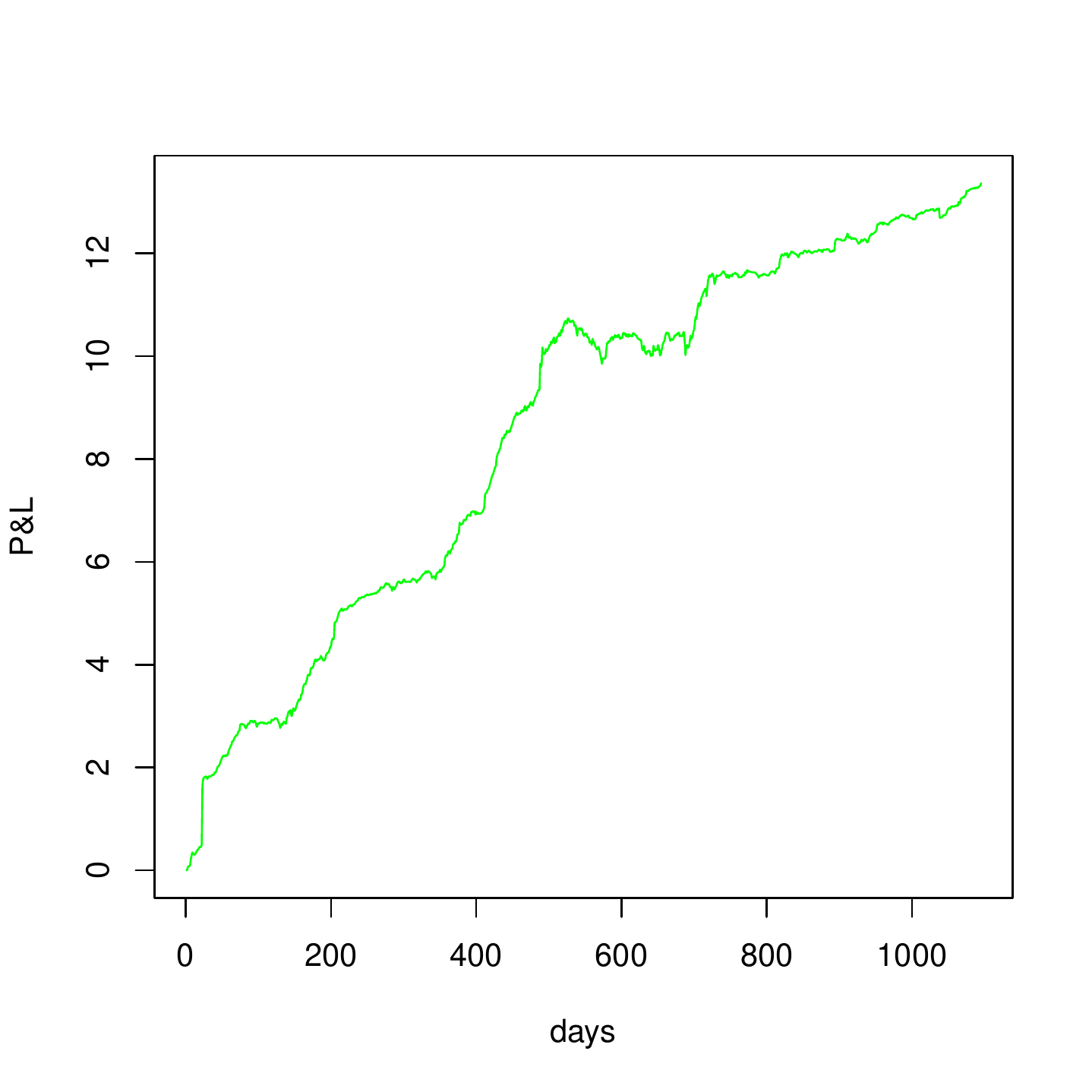}
\caption{P\&L (in the units where the short Bitcoin position is normalized to 1) for the lookback 3 portfolio (see Table \ref{table2}).}
\label{Figure8}
\end{figure}

\begin{figure}[ht]
\centering
\includegraphics[scale=1.0]{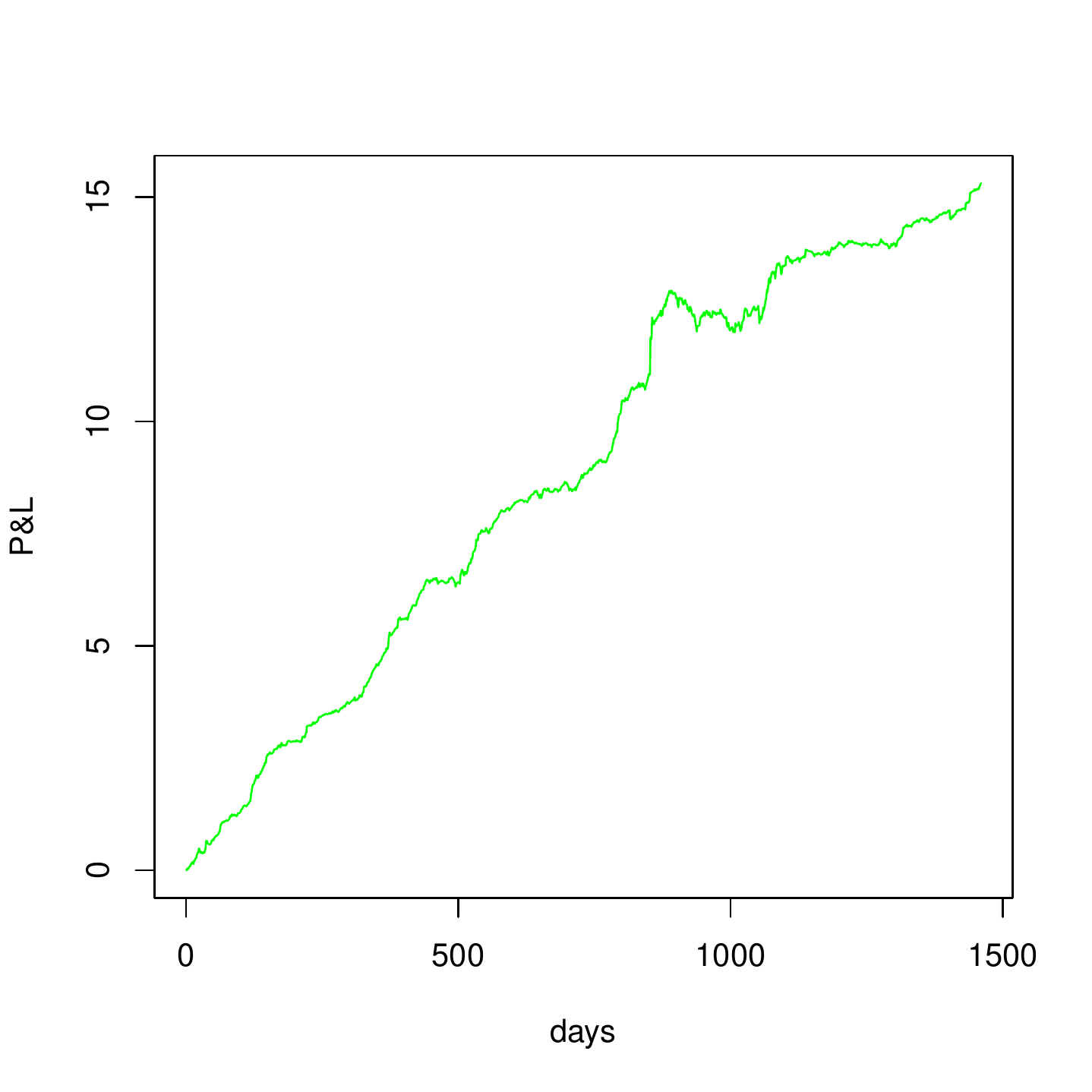}
\caption{P\&L (in the units where the short Bitcoin position is normalized to 1) for the lookback 4 portfolio (see Table \ref{table2}).}
\label{Figure9}
\end{figure}

\begin{figure}[ht]
\centering
\includegraphics[scale=1.0]{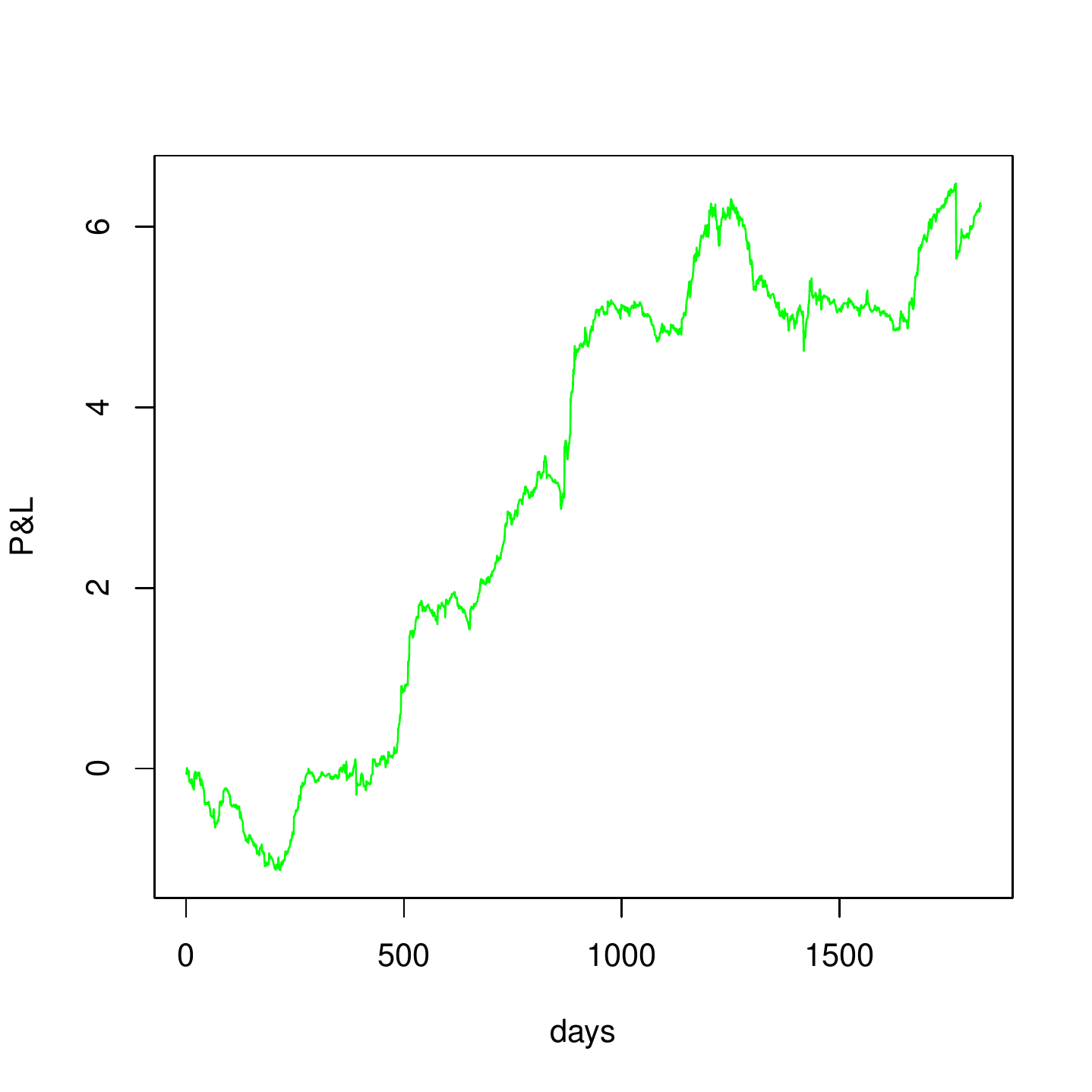}
\caption{P\&L (in the units where the short Bitcoin position is normalized to 1) for the lookback 5 portfolio (see Table \ref{table2}).}
\label{Figure10}
\end{figure}

{}For various universes mentioned above, Table \ref{table1} summarizes the market cap, the 20-day average daily volume, and the daily ``turnover" defined as the market cap divided by the 20-day average daily volume. Table 2 and Figures \ref{Figure1}-\ref{Figure10} summarize the backtest results. These results suggest that the altcoin-Bitcoin arbitrage alpha is a low-liquidity effect. It simply is not there for higher liquidity altcoins. However, the alpha -- which hinges on the mean-reversion effect based on the prior day's momentum (mom) -- is real. Thus, if for the universe 1E (see Tables \ref{table1} and \ref{table2}) we reverse the signal, we get ROC = -235.16\% and Sharpe = -7.18, and if we go long all altcoins with equal weights in this universe (irrespective of mom),\footnote{\, In this case we have a quasi-static (the altcoin universe can change with market cap fluctuations) dollar-neutral portfolio, which is obtained by setting $\alpha_{is}\equiv 1$ ($i=2,\dots,N$) in Eq. (\ref{equal}).} we get ROC = -26.56\% and Sharpe = -1.34. Conversely, if we reverse the signal for, say, the universe 1A, we do not get a positive return. Therefore, the altcoin-Bitcoin arbitrage alpha appears to be a real effect owing to low liquidity of the altcoins for which it is present. Put differently, the alpha exists as it cannot be arbitraged away.

\section{Concluding Remarks}\label{sec4}

{}So, the altcoin-Bitcoin arbitrage alpha we discuss above is essentially is ``a low-liquidity premium" in altcoin returns. In practice, to arbitrage it, one must account for trading costs -- both transaction costs and market impact. For low-liquidity altcoins the market impact can quickly become prohibitive when attempting to execute sizable trades. In fact, for the 2-year, 3-year, 4-year and 5-year lookbacks (where the number of altcoins with historical data is smaller) the na{\" i}ve pickup in the performance is due to the fact that most of the altcoins in these universes, even though they have been around for a while, are lower-cap, lower-liquidity cryptoassets (which is a telltale sign for persistence), with the notable exception of XRP (Ripple), which is what ``Max" in the market cap corresponds to in Table \ref{table1} for these universes.

{}Can the altcoin-Bitcoin arbitrage alpha -- or the momentum indicator on which it is based -- be useful outside outright arbitraging it (which may be challenging due to the liquidity considerations)? Perhaps. In a sideways cryptoasset market, this indicator could be used as a guide for executing trades for lower-liquidity altcoins in other contexts. Thus, statistically, we expect that there is a mean-reversion effect, and if its yesterday's momentum is positive, today an altcoin (on average) is expected to trade lower, and if its yesterday's momentum is negative, today said altcoin (on average) is expected to trade higher. This can then conceivably be used as a ``shorter-horizon" (daily) execution signal for longer-horizon trades. It should be mentioned, however, that the aforesaid alpha is a statistical effect, which is expected to work better for a sizable cross-section of altcoins, so using it as a ``shorter-horizon" execution signal for such sizable cross-sections of altcoins would make more sense than for a single (or a few) altcoin(s). In this regard, let us mention \cite{LiewLB2018}, whose conclusion is that forecasting (using machine learning techniques) short-horizon single-cryptoasset returns (for the top 100 cryptoassets by market cap) appears to be challenging. This also bodes well with our findings here.

\appendix
\section{R Source Code: Trading Signal}\label{app.A}

{}In this appendix we give R (R Project for Statistical Computing, \url{https://www.r-project.org/}) source code for computing the altcoin-Bitcoin arbitrage trading signal. The sole function {\tt{\small crypto.arb()}} reads the aggregated data files {\tt{\small cr.prc.txt}} (close price), {\tt{\small cr.open.txt}} (open price), {\tt{\small cr.high.txt}} (high price), {\tt{\small cr.low.txt}} (low price), {\tt{\small cr.vol.txt}} (dollar volume), {\tt{\small cr.cap.txt}} (market cap), {\tt{\small cr.name.txt}} (names of the cryptoassets in the same order as all the other files), and {\tt{\small cr.mnbl.txt}} (1 if the name is minable, otherwise 0) generated by the function {\tt{\small crypto.prc.files()}} of \cite{CryptoFM} (also, see fn.\ref{fn.data} hereof). Internally, {\tt{\small crypto.arb()}} computes the {\tt{\small av}} (average volume), {\tt{\small size}} (market cap), {\tt{\small mom}} (momentum), {\tt{\small hlv}} (intraday volatility) factors of \cite{CryptoFM} and the trading signal based on {\tt{\small mom}} (together with the trading universe based on {\tt{\small size}}). The inputs of {\tt{\small crypto.arb()}} are {\tt{\small days}} (the length of the selection period used in fixing the cryptoasset universe by applying the aforesaid non-NA data and non-zero volume filters, which period is further ``padded" -- see below), {\tt{\small back}} (the length of the skip period, i.e., how many days to skip in the selection period before the lookback period),\footnote{\, In our backtests we always set {\tt{\small back = 0}}. Also note that {\tt{\small mnbl}} and {\tt{\small av}} are not used internally.} {\tt{\small lookback}} (the length of the lookback period over which the backtest is run), {\tt{\small d.r}} (the extra ``padding" added to the selection period plus one day, so the moving averages can be computed out-of-sample; we take {\tt{\small d.r = 20}}), {\tt{\small d.v}} (the {\tt{\small av}} moving average length; we take {\tt{\small d.v = 20}}), {\tt{\small d.i}} (the {\tt{\small hlv}} moving average length; we take {\tt{\small d.i = 20}}), {\tt{\small ix.upper}} (the rank of the highest market cap (as of the previous day) altcoin to include in the trading universe), and {\tt{\small ix.lower}} (the rank of the lowest market cap (as of the previous day) altcoin to include in the trading universe). The function {\tt{\small crypto.arb()}} internally computes and plots/outputs the daily P\&L and annualized ROC and Sharpe ratio.\\
\\
{\tt{\small
\noindent crypto.arb <- function (days = 365, back = 0, lookback = days,\\
\indent d.r = 20, d.v = 20, d.i = 20, ix.lower = NA, ix.upper = 2)\\
\{\\
\indent read.prc <- function(file, header = F, make.numeric = T)\\
\indent \{\\
\indent \indent x <- read.delim(file, header = header)\\
\indent \indent x <- as.matrix(x)\\
\indent \indent if(make.numeric)\\
\indent \indent \indent mode(x) <- "numeric"\\
\indent \indent return(x)\\
\indent \}\\
\\
\indent calc.mv.avg <- function(x, days, d.r)\\
\indent \{\\
\indent \indent if(d.r == 1)\\
\indent \indent \indent return(x[, 1:days])\\
\indent \indent y <- matrix(0, nrow(x), days)\\
\indent \indent for(i in 1:days)\\
\indent \indent \indent y[, i] <- rowMeans(x[, i:(i + d.r - 1)], na.rm = T)\\
\\
\indent \indent return(y)\\
\indent \}\\
\\
\indent prc <- read.prc("cr.prc.txt")\\
\indent cap <- read.prc("cr.cap.txt")\\
\indent high <- read.prc("cr.high.txt")\\
\indent low <- read.prc("cr.low.txt")\\
\indent vol <- read.prc("cr.vol.txt")\\
\indent open <- read.prc("cr.open.txt")\\
\indent mnbl <- read.prc("cr.mnbl.txt")\\
\indent name <- read.prc("cr.name.txt", make.numeric = F)\\
\\
\indent d <- days + d.r + 1\\
\indent prc <- prc[, 1:d]\\
\indent cap <- cap[, 1:d]\\
\indent high <- high[, 1:d]\\
\indent low <- low[, 1:d]\\
\indent vol <- vol[, 1:d]\\
\indent open <- open[, 1:d]\\
\\
\indent take <- rowSums(is.na(prc)) == 0 \& rowSums(is.na(cap)) == 0 \&\\
\indent \indent rowSums(is.na(high)) == 0 \& rowSums(is.na(low)) == 0 \&\\
\indent \indent rowSums(is.na(vol)) == 0 \& rowSums(is.na(open)) == 0 \&\\
\indent \indent rowSums(vol == 0) == 0\\
\\
\indent ret <- log(prc[take, -d] / prc[take, -1])\\
\indent ret.d <- prc[take, -d] / prc[take, -1] - 1\\
\indent prc <- prc[take, -1]\\
\indent cap <- cap[take, -1]\\
\indent high <- high[take, -1]\\
\indent low <- low[take, -1]\\
\indent vol <- vol[take, -1]\\
\indent open <- open[take, -1]\\
\indent mnbl <- mnbl[take, 1]\\
\indent name <- name[take, 1]\\
\\
\indent if(back > 0)\\
\indent \{\\
\indent \indent ret <- ret[, (back + 1):ncol(ret)]\\
\indent \indent ret.d <- ret[, (back + 1):ncol(ret)]\\
\indent \indent prc <- prc[, (back + 1):ncol(prc)]\\
\indent \indent cap <- cap[, (back + 1):ncol(cap)]\\
\indent \indent high <- high[, (back + 1):ncol(high)]\\
\indent \indent low <- low[, (back + 1):ncol(low)]\\
\indent \indent vol <- vol[, (back + 1):ncol(vol)]\\
\indent \indent open <- open[, (back + 1):ncol(open)]\\
\indent \}\\
\indent days <- lookback\\
\\
\indent av <- log(calc.mv.avg(vol, days, d.v))\\
 \indent hlv <- (high - low)\^{}2 / prc\^{}2\\
\indent hlv <- 0.5 * log(calc.mv.avg(hlv, days, d.i))\\
\indent take <- rowSums(!is.finite(hlv)) == 0 \#\#\# removes stale prices\\
\\
\indent av <- av[take, ]\\
\indent hlv <- hlv[take, ]\\
\indent mom <- log(prc[take, 1:days] / prc[take, 1:days + 1])\\
\indent size <- log(cap)[take, 1:days]\\
\indent ret <- ret[take, 1:days]\\
\indent ret.d <- ret.d[take, 1:days]\\
\indent mnbl <- mnbl[take]\\
\indent name <- name[take]\\
\\
\indent pnl <- rep(0, days)\\
\indent for(i in days:1)\\
\indent \{\\
\indent \indent x <- -sign(mom[, i]) \#\#\# momentum based trading signal\\
\indent \indent \#\#\# x[] <- 1 \#\#\# no signal\\
\indent \indent \#\#\# x <- -x \#\#\# reverse signal\\
\\
\indent \indent sort.size <- sort(size[, i], decreasing = T)\\
\indent \indent if(is.na(ix.lower))\\
\indent \indent \indent ix.lower <- length(sort.size)\\
\indent \indent take <- size[, i] >= sort.size[ix.lower] \&\\
\indent \indent \indent size[, i] <= sort.size[ix.upper] \#\#\# cap tier based universe\\
\\
\indent \indent x[!take] <- 0\\
\indent \indent x[1] <- 0\\
\indent \indent if(days > 365)\\
\indent \indent \indent x[name == "Circuits of V..." ] <- 0 \#\#\# removes COVAL\\
\\
\indent \indent x[x < 0] <- 0\\
\indent \indent \#\#\# ss <- exp(hlv[, i]) \#\#\# hlv based volatility\\
\indent \indent \#\#\# ss can be computed e.g. as 20-day historical return volatility\\
\indent \indent \#\#\# x <- x / ss \#\#\# volatility-suppressed signal\\
\indent \indent \#\#\# x <- x * abs(mom[, i]) / ss\^{}2 \#\#\# alternative suppression\\
\indent \indent x <- x / sum(abs(x))\\
\indent \indent x <- x * ret.d[, i]\\
\indent \indent x <- sum(x) - ret.d[1, i]\\
\indent \indent if(!is.finite(x))\\
\indent \indent \indent x <- 0\\
\indent \indent pnl[i] <- x\\
\indent \}\\
\indent pnl1 <- pnl\\
\indent pnl <- pnl[days:1]\\
\indent pnl <- cumsum(pnl)\\
\indent plot(1:days, pnl, type = "l",\\
\indent \indent col = "green", xlab = "days", ylab = "P\&L")\\
\\
\indent roc <- round(mean(pnl1) * 365 / 2 * 100, 2) \#\#\# annualized ROC\\
\indent sr <- round(mean(pnl1) / sd(pnl1) * sqrt(365), 2) \#\#\# annualized Sharpe\\
\indent print(roc)\\
\indent print(sr)\\
\}
}}

\section{DISCLAIMERS}\label{app.B}

{}Wherever the context so requires, the masculine gender includes the feminine and/or neuter, and the singular form includes the plural and {\em vice versa}. The author of this paper (``Author") and his affiliates including without limitation Quantigic$^\circledR$ Solutions LLC (``Author's Affiliates" or ``his Affiliates") make no implied or express warranties or any other representations whatsoever, including without limitation implied warranties of merchantability and fitness for a particular purpose, in connection with or with regard to the content of this paper including without limitation any code or algorithms contained herein (``Content").

{}The reader may use the Content solely at his/her/its own risk and the reader shall have no claims whatsoever against the Author or his Affiliates and the Author and his Affiliates shall have no liability whatsoever to the reader or any third party whatsoever for any loss, expense, opportunity cost, damages or any other adverse effects whatsoever relating to or arising from the use of the Content by the reader including without any limitation whatsoever: any direct, indirect, incidental, special, consequential or any other damages incurred by the reader, however caused and under any theory of liability; any loss of profit (whether incurred directly or indirectly), any loss of goodwill or reputation, any loss of data suffered, cost of procurement of substitute goods or services, or any other tangible or intangible loss; any reliance placed by the reader on the completeness, accuracy or existence of the Content or any other effect of using the Content; and any and all other adversities or negative effects the reader might encounter in using the Content irrespective of whether the Author or his Affiliates is or are or should have been aware of such adversities or negative effects.

{}The R code included in Appendix \ref{app.A} hereof is part of the copyrighted R code of Quantigic$^\circledR$ Solutions LLC and is provided herein with the express permission of Quantigic$^\circledR$ Solutions LLC. The copyright owner retains all rights, title and interest in and to its copyrighted source code included in Appendix \ref{app.A} hereof and any and all copyrights therefor.


\newpage


\begin{table}[ht]
\caption{Summaries for market cap (Cap), 20-day average daily volume (ADV), and ``turnover" (Tvr) defined as Cap divided by ADV. All quantities are as of January 18, 2019. The suffix in the first column is the number of lookback years (and the Cap tier for the 1-year lookback -- see Subsection \ref{sub.univ}). Qu = quartile. See Table \ref{table2} for the numbers of altcoins in each trading universe. The 1A quantities include Bitcoin (along with 29 altcoins).} 
\begin{tabular}{l l l l l l l} 
\\
\hline\hline 
Quantity & Min & 1st Qu & Median & Mean & 3rd Qu & Max\\
Cap.1A & 1.19e+08 & 1.85e+08 & 5.13e+08 & 3.71e+09 & 1.86e+09 & 6.43e+10 \\
ADV.1A & 1.35e+06 & 9.90e+06 & 3.56e+07 & 5.03e+08 & 2.05e+08 & 5.18e+09 \\
Tvr.1A & 8.48e-03 & 2.92e-02 & 5.97e-02 & 1.93e-01 & 2.08e-01 & 1.93e+00 \\
Cap.1B & 3.75e+07 & 4.45e+07 & 5.77e+07 & 6.94e+07 & 9.85e+07 & 1.16e+08 \\
ADV.1B & 4.04e+04 & 6.59e+05 & 1.91e+06 & 3.76e+06 & 3.57e+06 & 2.08e+07 \\
Tvr.1B & 9.94e-04 & 9.90e-03 & 3.02e-02 & 5.15e-02 & 5.23e-02 & 3.01e-01 \\
Cap.1C & 1.66e+07 & 2.00e+07 & 2.42e+07 & 2.50e+07 & 2.81e+07 & 3.71e+07 \\
ADV.1C & 1.29e+03 & 3.22e+05 & 5.23e+05 & 1.90e+06 & 1.34e+06 & 1.75e+07 \\
Tvr.1C & 7.78e-05 & 1.20e-02 & 2.30e-02 & 6.80e-02 & 6.19e-02 & 5.85e-01 \\
Cap.1D & 4.38e+06 & 5.98e+06 & 7.62e+06 & 8.72e+06 & 1.10e+07 & 1.63e+07 \\
ADV.1D & 3.67e+02 & 4.10e+04 & 1.79e+05 & 7.94e+05 & 5.29e+05 & 3.14e+07 \\
Tvr.1D & 5.40e-05 & 6.03e-03 & 2.40e-02 & 1.12e-01 & 6.38e-02 & 4.83e+00 \\
Cap.1E & 1.02e+06 & 1.43e+06 & 2.20e+06 & 2.34e+06 & 2.96e+06 & 4.37e+06 \\
ADV.1E & 4.00e+02 & 3.60e+03 & 1.16e+04 & 1.14e+05 & 4.56e+04 & 3.04e+06 \\
Tvr.1E & 1.45e-04 & 2.00e-03 & 5.49e-03 & 4.56e-02 & 2.48e-02 & 1.02e+00 \\
Cap.1F & 8.82e+03 & 1.60e+05 & 4.22e+05 & 4.37e+05 & 7.15e+05 & 1.02e+06 \\
ADV.1F & 1.10e+01 & 4.97e+02 & 1.47e+03 & 4.50e+04 & 8.13e+03 & 3.54e+06 \\
Tvr.1F & 1.38e-04 & 1.87e-03 & 5.28e-03 & 1.09e-01 & 2.18e-02 & 9.38e+00 \\
Cap.2 & 2.44e+04 & 1.09e+06 & 4.70e+06 & 3.00e+08 & 2.49e+07 & 1.35e+10 \\
ADV.2 & 7.70e+01 & 4.30e+03 & 2.72e+04 & 3.75e+07 & 3.98e+05 & 2.75e+09 \\
Tvr.2 & 7.78e-05 & 2.34e-03 & 8.75e-03 & 3.24e-02 & 2.35e-02 & 5.03e-01 \\
Cap.3 & 6.70e+04 & 1.24e+06 & 5.07e+06 & 5.05e+08 & 2.21e+07 & 1.35e+10 \\
ADV.3 & 8.30e+01 & 7.06e+03 & 3.64e+04 & 6.14e+07 & 2.37e+05 & 2.75e+09 \\
Tvr.3 & 7.64e-04 & 2.33e-03 & 8.32e-03 & 2.67e-02 & 2.36e-02 & 2.75e-01 \\
Cap.4 & 6.70e+04 & 2.34e+06 & 8.30e+06 & 4.60e+08 & 3.62e+07 & 1.35e+10 \\
ADV.4 & 8.30e+01 & 8.88e+03 & 7.95e+04 & 2.99e+07 & 3.85e+05 & 5.27e+08 \\
Tvr.4 & 7.64e-04 & 2.25e-03 & 9.47e-03 & 2.72e-02 & 2.32e-02 & 2.75e-01 \\
Cap.5 & 1.31e+06 & 4.57e+06 & 1.27e+07 & 1.31e+09 & 8.23e+07 & 1.35e+10 \\
ADV.5 & 3.07e+03 & 1.88e+04 & 6.18e+04 & 8.67e+07 & 4.53e+06 & 5.27e+08 \\
Tvr.5 & 1.05e-03 & 2.33e-03 & 9.12e-03 & 3.64e-02 & 1.86e-02 & 2.75e-01 \\ [1ex] 
\hline 
\end{tabular}
\label{table1} 
\end{table}
\newpage
\begin{table}[ht]
\caption{Backtest results. Lookback is the number of years (and the Cap tier for the 1-year lookback -- see Subsection \ref{sub.univ}). The second column labeled ``rank" determines which altcoins are allowed to be included in the long portfolio on a given day: ``rank" refers to the range of the rank of the market cap on the previous day among all cryptoassets (with a few cryptoassets excluded as set forth in fn.\ref{fn.COVAL} hereof). ROC = annualized return-on-capital (which is given by $365 \times \mbox{P} / (I_L + I_S)$, where $P$ is the average daily P\&L, $I_L$ and $I_S$ are the total long and short investment levels, and $I_L = I_S$ for dollar-neutrality). Sharpe = annualized Sharpe ratio (which is given by $\sqrt{365} \times P / S$, where $S$ is the standard deviation of daily P\&Ls).} 
\begin{tabular}{l l l l} 
\\
\hline\hline 
Lookback & rank & ROC (\%) & Sharpe \\
1A & 2-30	& -44.25 & -1.81 \\
1B & 31-60 & -59.29 & -1.99 \\
1C & 61-100 & -24.78 & -0.83 \\
1D & 101-200 & 11.26 & 0.43 \\
1E & 201-300 & 124.43 & 4.93 \\
1F & 301-417 & 459.08 & 14.65 \\
2 & 2-121 & 174.7 & 4.2 \\
3 & 2-67 & 222.75 & 4.06 \\
4 & 2-44 & 191.43 & 4.34 \\
5 & 2-13 & 62.27 & 1.28\\ [1ex] 
\hline 
\end{tabular}
\label{table2} 
\end{table}
%


\newpage\clearpage

\end{document}